\documentstyle[twocolumn,prl,aps,ams,epsf]{revtex}
\begin{document}
\draft
\title{On the Mechanism of Time--Delayed Feedback Control}
\author{Wolfram Just\thanks{e--mail: 
wolfram@arnold.fkp.physik.th-darmstadt.de}, 
Thomas Bernard, Matthias Ostheimer, 
Ekkehard Reibold\thanks{e--mail: 
reibold@exp1.fkp.physik.th-darmstadt.de}, 
and Hartmut Benner\thanks{e--mail: 
benner@hrzpub.th-darmstadt.de}}
\address{Institut f\"ur Festk\"orperphysik, Technische Hochschule Darmstadt,
Hochschulstra\ss e 6, D--64289  Darmstadt, Germany}
\date{November 15, 1996}
\maketitle
\begin{abstract}
The Pyragas method for controlling chaos 
is investigated in detail from the experimental as well as 
theoretical point of view. We show by an analytical stability analysis 
that the revolution around an unstable periodic orbit governs the success 
of the control scheme. Our predictions concerning the transient 
behaviour of the control signal are confirmed by numerical simulations and 
an electronic circuit experiment.
\end{abstract}
\newcommand{\be}{\begin{equation}\label}
\newcommand{\ee}{\end{equation}}
\newcommand{\bea}{\begin{eqnarray}\label}
\newcommand{\eea}{\end{eqnarray}}
\newcommand{\vx}{\mbox{\boldmath$\!\!x$}}
\newcommand{\vf}{\mbox{\boldmath$\!\!f$}}
\newcommand{\vF}{\mbox{\boldmath$\!\!F$}}
\newcommand{\vz}{\mbox{\boldmath$\!\!z$}}
\newcommand{\vu}{\mbox{\boldmath$\!\!u$}}
\newcommand{\vv}{\mbox{\boldmath$\!\!v$}}
\newcommand{\vxi}{\mbox{\boldmath$\!\!\xi$}}
\section{Introduction}
The problem of controlling unstable motion is a
classical subject in engineering science. 
The revived interest of physicists in this subject, however, started 
with the observation that a large number of
unstable periodic orbits embedded in chaotic attractors
can be stabilized by weak external forces \cite{ogy}. Since that
time a real industry on chaos control has developed \cite{ind}.
Two main methods for controlling unstable motions have been
established meanwhile. The first one, 
developed by Ott, Grebogi, and Yorke \cite{ogy}, is based
on the invariant mani\-fold structure of unstable orbits.
It is theoretically well understood, but  difficult 
to apply to fast experimental systems. The second one,
proposed by Pyragas \cite{Pyragas}, uses time--delayed 
controlling forces. In contrast
to the former one it can easily be applied to real experimental situations,
but so far the control mechanism has been poorly understood from a theoretical
point of view. 
By performing an analytical linear stability analysis we
demonstrate which class of orbits is accessible to
time--delayed feedback control methods. In addition, we obtain
explicit expressions for important quantities like the critical and
optimal control amplitude or the dependence of the transient
behaviour on the control parameters.

\section{Theoretical Approach}
\begin{figure}
\epsffile{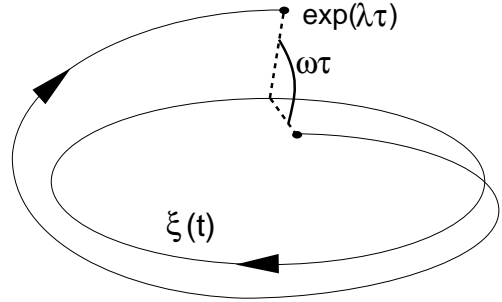} \vspace*{3mm}
\caption{Diagrammatic view of a trajectory in the vicinity of an
unstable periodic orbit.}
\label{figa}
\end{figure}

We consider a dynamical system which is described by a general set
of differential equations. It may contain a periodic explicit time 
dependence. 
\be{aa}
\dot{\vx}= \vf\left(\vx(t),t\right)
\ee
We are interested in the stabilization of an unstable periodic orbit
$\vxi(t)=\vxi(t+\tau)$. 
$\tau$ is an integer multiple of the period of the driving force
for non--autonomous systems. 
We remind the reader that the linear
stability analysis of such an orbit
according to $\vx(t)=\vxi(t)+\exp\left[(\lambda+i\omega)t\right]\vu(t)$
leads to a Floquet problem, where the exponent and the periodic
eigenfunction $\vu(t)=\vu(t+\tau)$ are 
determined by
\be{ab}
\left[\lambda+i\omega\right] \vu(t) + \dot{\vu} = 
D \vf(\vxi(t),t) \vu(t) \quad ,
\ee
with $D\vf$ denoting the Jacobian matrix of $\vf$. 
Since our subsequent analysis 
applies separately 
to each Floquet exponent
we refrain from numbering the different branches.
The real and the imaginary parts of the Floquet exponents govern the
instability and the revolution of the trajectory
around the unstable periodic orbit
(cf. fig.\ref{figa}).

In order to achieve control of the unstable periodic orbit the system 
(\ref{aa}) is, following the idea of \cite{Pyragas}, subjected to a
time--delayed force $K\left[\vx(t)-\vx(t-\tau)\right]$. 
The most general situation is given by
\be{ac}
\dot{\vz}= \vF\left(\vz(t),K\left[\vz(t)-\vz(t-\tau)\right],t\right)
\quad ,
\ee
where the right--hand side obeys the constraint
$\vF(\vz,0,t)=\vf(\vz,t)$, and the amplitude $K$ of the controlling
force is introduced for convenience. As long as the delay time
coincides with the period of the unstable periodic orbit
the controlled system admits the same solution $\vz(t)=\vxi(t)$.
Linear stability analysis according to $\vz(t)=\vxi(t)+\delta \vz(t)$
yields
\bea{ad}
\delta \dot{\vz} &=& D_1 \vF\left(\vxi(t),0,t\right) \delta \vz(t)\nonumber\\
&+& 
D_2 \vF\left(\vxi(t),0,t\right) K \left[\delta \vz(t)- \delta \vz(t-\tau)
\right]\quad ,
\eea
where $D_i \vF$ denotes the Jacobian matrix
with respect to the $i^{th}$ (vector type) argument.
In the case of conventional Pyragas control, where only one
system variable is assumed to be accessible, 
the matrix $D_2 \vF$
contains only one non--vanishing element on the diagonal.
But we keep our approach as general as possible.
The (infinite dimensional generalization of) 
Floquet theory \cite{Hale}
tells us that the deviations obey 
$\delta \vz (t)= \exp\left[(\Lambda + i \Omega) t \right] \vv(t)$
and $\vv(t)=\vv(t+\tau)$, so that eq.(\ref{ad}) reduces to
\bea{ae}
& &\left[\Lambda+i\Omega\right] \vv(t) + \dot{\vv}\nonumber\\ 
&=& 
A\left[K \left( 1- \exp\left[-\Lambda \tau - i\Omega \tau \right]
\right) ,t \right] \vv(t) \quad . 
\eea
Here the abbreviation
\be{af}
A\left[\kappa,t\right] := D \vf(\vxi(t),t) + D_2 \vF
\left(\vxi(t),0,t\right) \kappa
\ee
has been used. From eq.(\ref{ae}) it is obvious that $\Lambda+i\Omega$
can be expressed in terms of the Floquet exponents of the matrix
(\ref{af}). If we denote the latter for convenience by
$\Gamma\left[\kappa\right]$ 
then eq.(\ref{ae}) implies the relation
\be{ag}
\Lambda + i \Omega = \Gamma \left[
K \left( 1- \exp\left[-\Lambda \tau -i\Omega \tau \right] \right) \right]
\quad .
\ee
This expression, which in fact is not entirely new but has been
evaluated only numerically for specific examples (cf.~\cite{Pyragas2,Verg}),
determines the exponents of the controlled orbit
in dependence of the control amplitude $K$.
Although it is in general a difficult task to obtain a closed analytical
expression for the quantity $\Gamma$, we know by definition 
that the boundary condition (cf.~eqs.(\ref{ab}) and (\ref{af}))
\be{ah}
\Gamma[0]=\lambda+i \omega
\ee
holds, and that $\Gamma$ is 
an analytical function as long as the Floquet exponents are non--degenerate.
These properties are sufficient to conclude that only orbits with a
finite frequency $\Omega\neq 0$ can become stable. On increasing
the control amplitude $K$ the real part of the Floquet exponent
$\Lambda$ has to change its sign from positive to negative values in
order to achieve stabilization. But if the frequency $\Omega$ of the
controlled orbit remains zero, the influence of the controlling force
$K[1-\exp(-\Lambda\tau)]$ vanishes if the orbit tends to become
stable, so that the solutions $\Lambda$ 
of eq.(\ref{ag}) never can change their sign. 
The reader might object that the condiditon $\Omega=0$ is atypical and
does not occur generically. But we remind of the fact that
(nondegenerate) real Floquet multipliers are stable with respect to
perturbations (cf.~\cite{GuHo}), so that both cases $\Omega=0$ and
$\Omega\neq 0$ occur in a sense with equal proability. However
one should keep in mind that the neccessary condition 
$\Omega\neq 0$ for stabilization has to be fulfilled for each Floquet branch
separately. Hence stabilization may be unlikely if the unstable manifold
is high--dimensional.

We summarize that
{\it by the Pyragas method only orbits with a finite torsion can be
stabilized}, since for stabilization
the influence of the controlling force 
has to be finite (cf.~fig.\ref{figa}).
This property has been observed recently even in high--dimensional
dynamical systems by analysing the transient behaviour of the
control signal \cite{hess}, but no explanation has been proposed.  
From this point of view the control methods
by Ott, Grebogi, and Yorke on the one hand and by 
Pyragas on the other hand are
complementary, since
the former is not in principle but in most practical applications 
restricted to the  case of only one unstable eigendirection, whereas
the latter requires at least
a two--dimensional unstable manifold. 

For further quantitative
investigations some information about $\Gamma[\kappa]$ is
required. There are a few cases where by inspection the 
$\kappa$--dependence can be read off from eq.(\ref{af}):
\bea{tmp}
D_2 \vF\left(\vxi(t),0,t\right) &=& 1 \nonumber\\
&\Rightarrow& \Gamma[\kappa] = \lambda + i\omega + \kappa
\label{aia}\\[2mm]
D_2 \vF\left(\vxi(t),0,t\right) &=&  D \vf(\vxi(t),t) \nonumber\\
&\Rightarrow& \Gamma[\kappa] = (\lambda + i \omega) (1+\kappa)
\label{aib}
\eea
But we do not intend to confine our analysis to these special
situations (cf.~\cite{bnrconf}). Instead we suppose that the
controlling force is small enough in order to neglect 
higher order terms in the expansion of eq.(\ref{ag}). 
\bea{aj}
& &\Lambda + i \Omega = \lambda + i \omega\nonumber\\ 
&+& (\chi'+i\chi'')
K \left( 1- \exp\left[-\Lambda\tau - i\Omega \tau\right] \right)
+ {\cal O}(K^2)
\eea
Here use has been made of relation (\ref{ah}), and the 
abbreviation $\chi'+i\chi'':= d \Gamma/d\kappa|_{\kappa=0}$ 
contains the details of the coupling mechanism
of the controlling force. It is worth to mention that
relation (\ref{aj}) is exactly valid for the cases described by 
eqs.(\ref{aia}), (\ref{aib}).

Relation (\ref{aj}) determines the stability of the controlled orbit
in terms of the control amplitude $K$, the Floquet exponent of the
uncontrolled orbit $\lambda+i\omega$, and the precise mechanism of the
coupling $\chi',\chi''$.
\bea{ak}
\Lambda &=& \lambda + K \chi'\left(1-\exp(-\Lambda\tau)\cos(\Omega \tau)
\right)\nonumber\\
&-& K \chi'' \exp(-\Lambda\tau)\sin(\Omega \tau) \label{aka}\\
\Omega &=& \omega + K \chi'\exp(- \Lambda \tau)\sin(\Omega \tau)\nonumber\\
&+&
K \chi'' \left(1-\exp(-\Lambda \tau) \cos(\Omega \tau)\right)
\label{akb}
\eea
For the evaluation we confine the subsequent discussion to an 
uncontrolled unstable periodic orbit which just flips its neighbourhood 
within one period, that means to an orbit of frequency $\omega=\pi/\tau$.
Such a situation appears particularly in a
neighbourhood of a period doubling bifurcation. Since the corresponding
Floquet exponent $\lambda+i\omega$ is located at the ''boundary of the
Brillouin zone'', that means the corresponding multiplier is an isolated
negative real number, $\Gamma[\kappa]-i \omega$ is by definition 
a real function at $\kappa=0$ (cf.~eq.(\ref{af})), and 
$\chi''$ vanishes. 
For this reason $\Omega=\pi/\tau$ is a
solution of eq.(\ref{akb}) which in that sense is an optimal value for
the frequency since the effect of the control term in eq.(\ref{aka})
becomes maximal. Then eq.(\ref{aka}) simplifies to
\be{al}
\Lambda = \lambda + K \chi' \left(1 + \exp(-\Lambda \tau ) \right), \quad
\Omega=\pi/\tau \quad .
\ee
Stabilization is achieved if $\Lambda$ changes its sign, that means at
a critical control amplitude
\be{am}
K_{c}= - \lambda/(2 \chi')\quad .
\ee
$K_c$ is mainly determined by the instability of the uncontrolled orbit.
On further increase of the control amplitude the frequency may start to
deviate from its optimal value. Formally this deviation results
from a pitchfork bifurcation in eq.(\ref{akb}) which occurs at
$K_{opt}$
\be{an}
1=- K_{opt} \chi' \tau \exp(- \Lambda_{opt} \tau) \quad ,
\ee
with $\Lambda_{opt}$ being determined by eq.(\ref{al}). Beyond $K_{opt}$
the frequency $\Omega$ deviates from its optimal value $\pi/\tau$ so that
the eigenvalue $\Lambda$ of the controlled orbit starts to increase 
again with
the control amplitude $K$. In this sense $K_{opt}$ is the optimal
value since the stability of the controlled orbit is maximal.

\section{Simulations and Experiment}
\begin{figure}
\epsffile{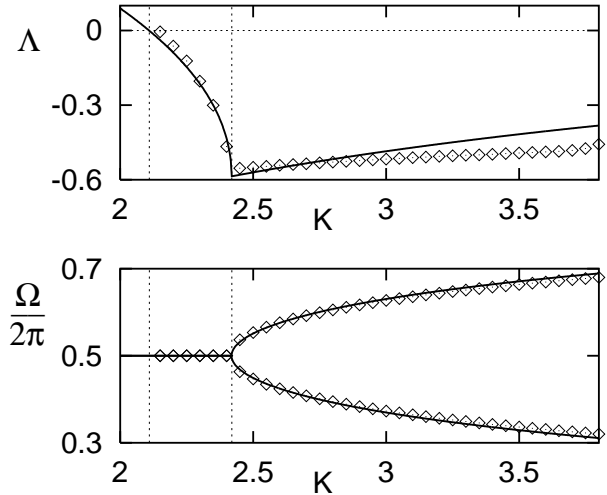} \vspace*{3mm}
\caption[ ]{Real and imaginary part of the Floquet exponent for a
numerical simulation of eqs.(\ref{aob}). The data
(symbols) have been obtained from
the decay of the control signal on variation of the
control amplitude $K$.
The solid lines indicate the analytical solutions of
eqs.(\ref{aka}) and (\ref{akb}) with
$\lambda\approx 0.97$, $\chi'\approx -0.23$.}
\label{figb}
\end{figure}

In order to illustrate our theoretical considerations and to demonstrate
that the features predicted are accessible from observed data, we
have performed numerical simulations of the driven
Toda oscillator
\begin{eqnarray}\label{ao}
\dot{z}_1 &=& z_2 \nonumber \\ 
\dot{z}_2 &=& - \mu z_2 - \alpha \left( \exp(z_1)-1\right) + 
A \sin 2\pi t \nonumber\\
&-& K \left[ z_2(t)-z_2 (t-1) \right] \label{aob}\quad .
\end{eqnarray}
Time is measured in periods of the driving force.
At $\mu=0.8$, $\alpha=25$, $A=105$, and $K=0$ 
the system possesses a chaotic attractor. 
A period--one orbit, which has become unstable in a
period doubling bifurcation, can be stabilized at finite control
$K>K_{c}\approx 2.1$. From the exponential decay of the control
signal $z_2(t)-z_2(t-1)$, using a standard least square method,  
we determine the real and imaginary parts of
the Floquet exponent $\Lambda+i\Omega$ for several
values of the control amplitude. Our findings are
summarized in fig.\ref{figb}. 
We clearly observe the predicted dependence on the control amplitude.
Using the value $K_{opt}\approx 2.4$ we determine the two parameters
$\lambda$ and $\chi'$ by means of eqs.(\ref{am}) and (\ref{an})
and compare our simulations with the theoretical prediction from
eqs.(\ref{aka}) and (\ref{akb}). The quantitative coincidence
is within a few percent.
This result is even more convincing when keeping in mind that
the theoretical prediction is just a first order computation in the 
control amplitude which actually is not so small. 
For larger values of $K$ we observe an additional frequency in
the control signal. Such a property can be attributed to the
second (stable) Floquet branch and may also be evaluated by eq.(\ref{ag}).
Finally, as a by--product,  
we obtain an estimate of the Floquet exponent  
$\lambda+i\omega$ of the uncontrolled orbit.

\begin{figure}
\epsffile{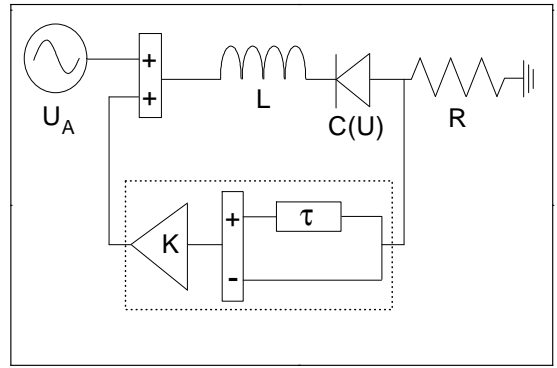} \vspace*{3mm}
\caption[ ]{Experimental setup of the nonlinear diode resonator
with time--delayed feedback device.}
\label{figd}
\end{figure}

\begin{figure}
\epsffile{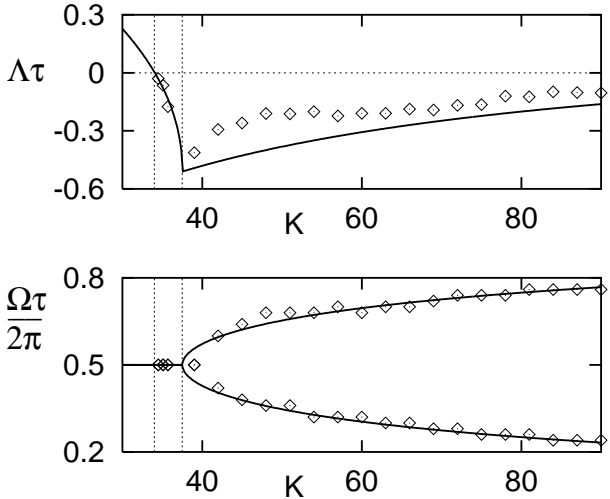} \vspace*{3mm}
\caption[ ]{Real and imaginary part of the Floquet exponent 
for the electronic circuit experiment. The data (symbols) have
been obtained from
the decay of the control signal on variation of the
control amplitude $K$.
The solid lines indicate the analytical solutions of
eqs.(\ref{aka}) and (\ref{akb}) with
$\lambda\tau \approx 1.09$, $\chi'\tau \approx -0.016$.}
\label{figc}
\end{figure}

In addition to 
analytical calculations and computer simulations we have performed
experiments on a nonlinear electronic circuit (see e.g.~\cite{Car}).
We consider a nonlinear diode resonator consisting of a 
capacity diode (1N4005),
an inductor ($470\mu \mbox{H}$), and a resistor ($40 \Omega$) 
(cf.~fig.\ref{figd}). 
The control device consists of a cascade of electronic delay lines with a
limiting frequency of about $2 \mbox{MHz}$ 
and several operational amplifiers acting
as preamplifier, subtractor, or inverter. 
The device allows to apply a
controlling force of the form $\pm K [U(t) - \epsilon U(t- \tau )] +
U_0$ 
with parameter ranges $K=0 \ldots 300$, $\epsilon = 0\ldots 2$, $\tau =
10\ldots 7000 \mbox{ns}$, 
and $U_0= -5\ldots +5\mbox{V}$. 
For conventional delayed feedback
control $\epsilon$ has carefully to be adjusted to one 
and the offset $U_0$ to zero. 
This was done in the experiment reported here.
The circuit was sinusoidally driven 
with an amplitude of $2\mbox{V}$ and
a frequency of $990\mbox{kHz}$. 
Accordingly the delay time was set to $\tau =
1010\mbox{ns}$.
From the transient dynamics of the control signal we again
obtain the decay rate and the frequency (cf.~fig.\ref{figc}).
For comparison with the theoretical prediction we
take the values $K_c\approx 34$ and $K_{opt}\approx 37.5$ to
determine the two parameters $\chi'\tau$ and
$\lambda \tau$ from eqs.(\ref{am}) and (\ref{an}).
The quantitative agreement with eqs.(\ref{aka}), (\ref{akb}) is
within a few percent, except for the real part beyond 
$K_{opt}$. 
Apart from the reasons already mentioned the
deviations can be attributed to the limited accuracy of
the value $K_{opt}$. Since the transients are affected by
noise a precise estimate of the exponents
is difficult to obtain for small decay rate $\Lambda$. 

\section{Conclusion}
We have shown that the main limiting factor for time--delayed feedback
control results from the torsion of the unstable periodic orbit.
This topological property determines
whether the control mechanism works at all. We have
worked out the general features of the transient behaviour
including critical and optimal control amplitudes. 
Our approach describes at least the generic properties
for stabilizing unstable periodic orbits with an unstable manifold
like a M\"obius strip.
Our simulations have shown that the features described
above are accessible from the transient behaviour of the control signal
and hence are observable in experiments. The electronic circuit
experiment demonstrates that an analysis along these lines is possible
even for ultrafast experiments. 
Our theoretical approach, based on an expansion of the general 
expression (\ref{ag}), resembles a Ginzburg--Landau like treatment 
of phase transitions. It can
be easily extended to incorporate e.g.~the degeneracy
of several Floquet exponents or the features of spatially extended,
that means high--dimensional systems.

\section{Acknowledgement}
This project of SFB 185 ''Nichtlineare Dynamik''  
was partly financed 
by special funds of the Deutsche Forschungsgemeinschaft.

\end{document}